\documentclass[]{spie}  
\usepackage[]{graphicx}

\title{Monte Carlo Simulation of HERD Calorimeter}

\author{M. Xu\supit{a}, G.M. Chen\supit{a}, Y.W. Dong\supit{a}, J.G. Lu\supit{b}, Z. Quan\supit{a},L. Wang\supit{c}, Z.G. Wang\supit{b}, B.B. Wu\supit{a}, S.N. Zhang\supit{a}
\skiplinehalf
\supit{a}Key Laboratory of Particle Astrophysics, Institute of High Energy Physics, Chinese Academy of Sciences, Beijing, China; \\
\supit{b}Center of Experimental Physics, Institute of High Energy Physics, Chinese Academy of Sciences, Beijing, China; \\
\supit{c}Xi'an Institute of Optics and Precision Mechanics, Chinese Academy of Sciences, Xi'an, China
}

\authorinfo{Further author information: (Send correspondence to M. Xu)\\Ming XU: E-mail: mingxu@ihep.ac.cn\\ }

\begin{document}
  \maketitle

\begin{abstract}
The High Energy cosmic-Radiation Detection (HERD) facility onboard China's Space Station is planned for operation starting around 2020 for about 10 years. It is designed as a next generation space facility focused on indirect dark matter search, precise cosmic ray spectrum and composition measurements up to the knee energy, and high energy gamma-ray monitoring and survey. The calorimeter plays an essential role in the main scientific objectives of HERD. A 3-D cubic calorimeter filled with high granularity crystals as active material is a very promising choice for the calorimeter. HERD is mainly composed of a 3-D calorimeter (CALO) surrounded by silicon trackers (TK) from all five sides except the bottom. CALO is made of 9261 cubes of LYSO crystals, corresponding to about 55 radiation lengths and 3 nuclear interaction lengths, respectively. Here the simulation results of the performance of CALO with GEANT4 and FLUKA are presented: 1) the total absorption CALO and its absorption depth for precise energy measurements (energy resolution: 1\%  for electrons and gamma-rays beyond 100 GeV, 20\% for protons from 100 GeV to 1 PeV); 2) its granularity for particle identification (electron/proton separation power better than $10^{-5}$); 3) the homogenous geometry for detecting particles arriving from every unblocked direction for large effective geometrical factor ($>$3 ${\rm m}^{2}{\rm sr}$ for electron and diffuse gamma-rays, $>$2 $ {\rm m}^{2}{\rm sr}$ for cosmic ray nuclei); 4) expected observational results such as gamma-ray line spectrum from dark matter annihilation and spectrum measurement of various cosmic ray chemical components.
\end{abstract}



\section{INTRODUCTION}
\label{sec:intro}  

The intensity of primary cosmic rays (CRs) in the energy range from several GeV to somewhat beyond 100 TeV follows a power law broken around several PeV (the knee). The knee refers to the steepening power law above several PeV. The origin of the knee has become a classic problem in CR physics since its discovery in 1958, as it is related closely to the physics of acceleration and propagation of CRs. Numerous mechanisms have been proposed to explain the turning shape in the CR all particle spectrum\cite{cr_model}, including the limitation of acceleration processes, particle leakage from the Galaxy, etc. Ground-based experiments\cite{kascade,tibet,argo} that rely on extensive air shower have made observations around the knee region\cite{ground_review}; however, due to the disadvantages of energy reconstruction and particle identification, these experiments can hardly reduce the energy uncertainty in the measurements and reveal the fine structure of the knee region. On the other hand, experiments based on balloons,\cite{atic2,cream1} satellites\cite{pamela} or the international space station\cite{ams02} can measure the particle energy directly and identify particles by charge with ionization measurement and other techniques. Unfortunately, due to the weight constraint and power consumption limitation in space, these experiments suffer from small geometrical factor and limited interaction lengths at higher energy. Because the number of CR particles falls rapidly as energy increases, to date there is no direct observation in space around the knee region.

The existence of dark matter (DM) in the universe is widely accepted. A variety of astrophysical observations indicate that the DM particle should be neutral, cold and non-baryonic. Weakly Interacting Massive Particles\cite{wimp} (WIMPs) form a particularly interesting generic class of new-particle candidates because they can account for the observed DM density naturally. WIMPs can be detected in CRs through its annihilation into electrons or gamma-rays, resulting in structures to be seen in the otherwise predicted smooth spectra. Anomalies on electron spectrum\cite{atic_electron,Fermi_electron} and additional positron fraction in CR\cite{pamela_positron,ams02_positron} have been observed. DM is an attractive interpretation; however, astrophysical sources like pulsars and pulsar wind nebulae can also contribute to these observed results. Experimental data from more precise measurement at higher energies are needed to the search of DM.

HERD onboard China's Space Station is planned for operation starting around 2020 for about 10 years. HERD is therefore focused on enlarging the detection geometrical factor to guarantee significant amount of data in the knee region with precise measurement. With the breakthrough on geometrical factor, HERD could also obtain discoveries from the possible structures of high energy electron and gamma-ray. Signals from DM particles and/or other new physics may show up in these observation channels. Please refer to Zhang et al (2014) in the same proceedings for futher details of HERD scientific objectives and requirements\cite{zhangsn}.
\section{Baseline Design of HERD}

\begin{figure}
\begin{center}
\begin{tabular}{c}
\includegraphics[height=7cm]{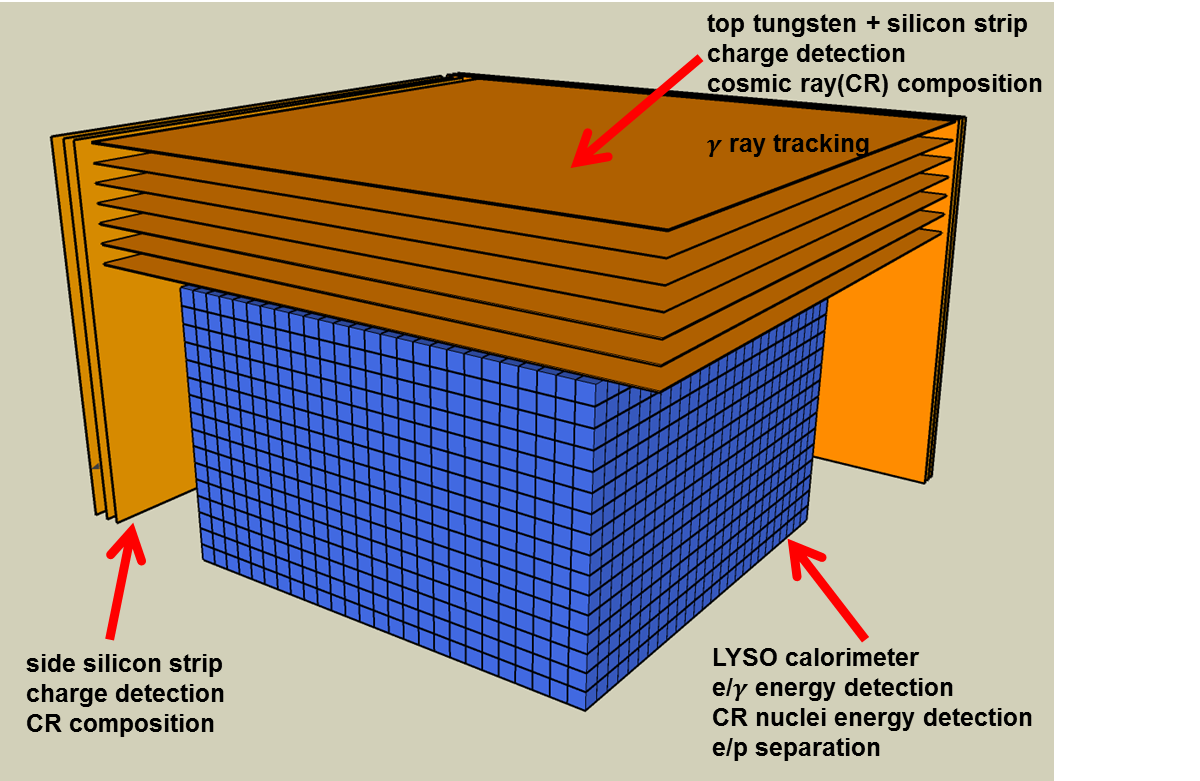}
\end{tabular}
\end{center}
\caption[example]
{ \label{fig:baseline}
Schematic diagram of HERD.}
\end{figure}

The baseline design of HERD includes two main parts, as shown in Fig.~\ref{fig:baseline}: the micro silicon strip tracker covering five sides except the bottom for mechanical support of the instrument, for particle trajectory and nuclei charge measurement; the LYSO crystal calorimeter, for particle energy and identification measurement.

The top tracker is constituted of micro silicon strips and thin tungsten foils, the latter act as converter for gamma-rays. A gamma-ray continues until it interacts with the thin tungsten foils, producing electron/positron pairs, and the primary direction can be reconstructed from the tracks of charged particles. There are seven tracking planes of x-y layers of silicon strips, each with the area of 65 $\times$ 65 ${\rm cm}^{2}$, and five converter layers of tungsten sandwiched between layers 2-6, with the thickness corresponding to two radiation length. The first layer of silicon can be used to determine if the incident particle is charged or not. The top tracker provides charge identification, trajectory measurement, back scatter rejection and some early shower development of gamma-rays and electrons. The other four sides in the baseline design are covered with only three x-y layers of silicon strip without tungsten foils, for nuclei charge and trajectory measurement; we are also studying the option of having the side trackers identical to the top tracker, in order to increase significantly the acceptance of gamma-rays.

CALO is composed of $21\times21\times21$ cubic LYSO crystals, with each cell of $3\times3\times3$ ${\rm cm}^{3}$ coupled with wavelength shifter fiber readout by ICCD. CALO corresponds to about 55 radiation lengths and 3 nuclear interaction lengths respectively, making it the deepest calorimeter ever used in space. The full active CALO obtains either the total energy deposition of electromagnetic shower induced by gamma-rays and electrons or at least partial energy deposition of hadronic shower induced by CR nuclei. Fine shower shape information will be given by its high granularity. The almost homogenous geometry for detecting particles arriving from every unblocked direction increases its geometrical factor under the same weight constraint.

\section{CALO Simulation}
\label{sec:simu}
Monte Carlo simulation for the CALO has been set up under the framework of GEANT4\cite{geant4} and FLUKA\cite{fluka_1} at the level of energy deposition in crystals, without modeling the readout chain response at the current stage. The simulated samples have been produced by using the interaction model QGSP for electron and gamma-ray incident and DPMJET-III for high energy proton case.

\subsection{Energy Measurement}
\label{sec:reso}
\begin{figure}
\begin{center}
\begin{tabular}{c}
\includegraphics[height=7cm]{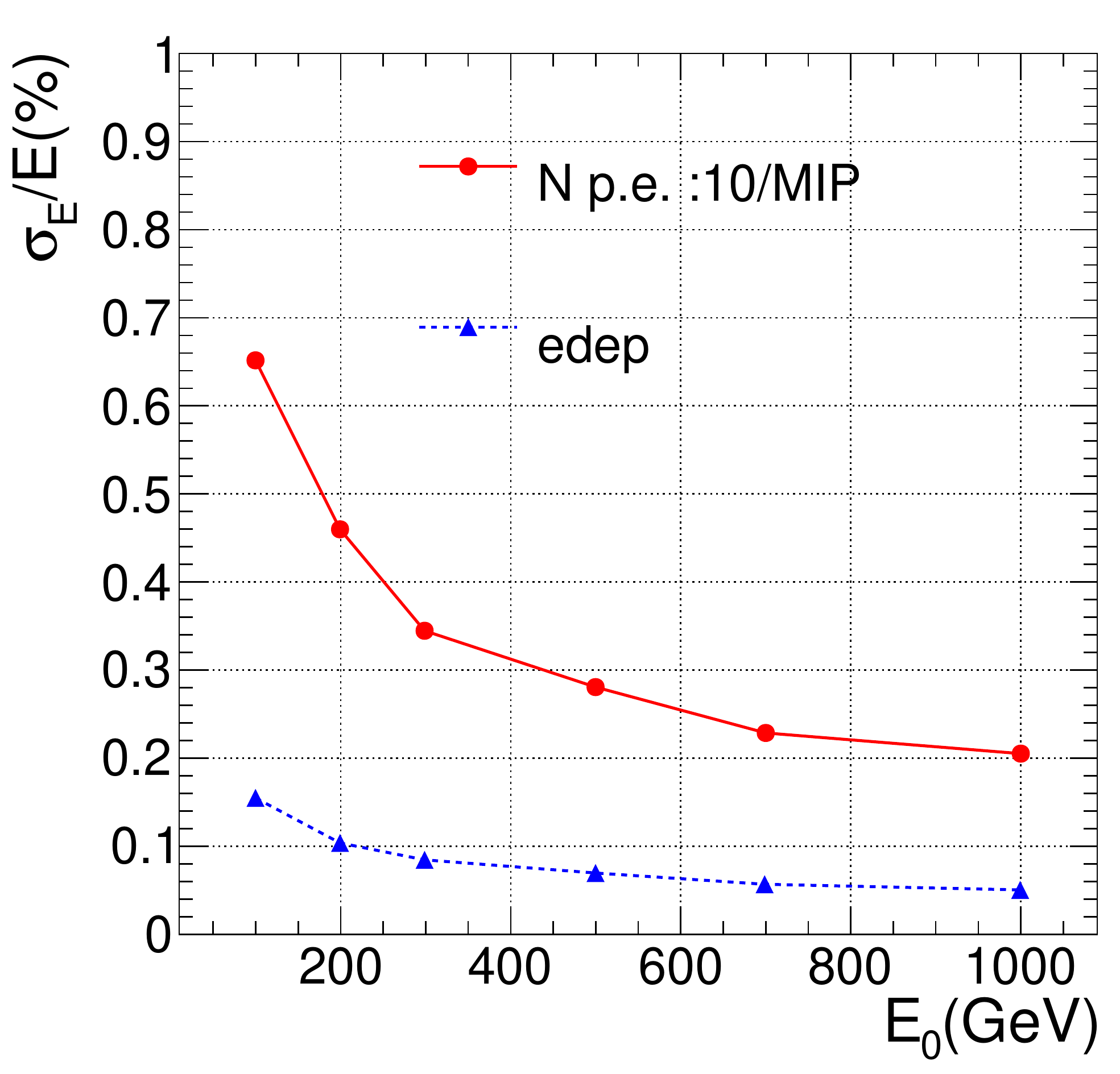}
\end{tabular}
\end{center}
\caption[example]
{ \label{fig:e_reso}
Energy resolution for electrons and gamma-rays. The dashed line is the theoretical energy resolution and the solid line takes into account the stochastic uncertainty.}
\end{figure}

\begin{figure}
\begin{center}
\begin{tabular}{c}
\includegraphics[height=7cm]{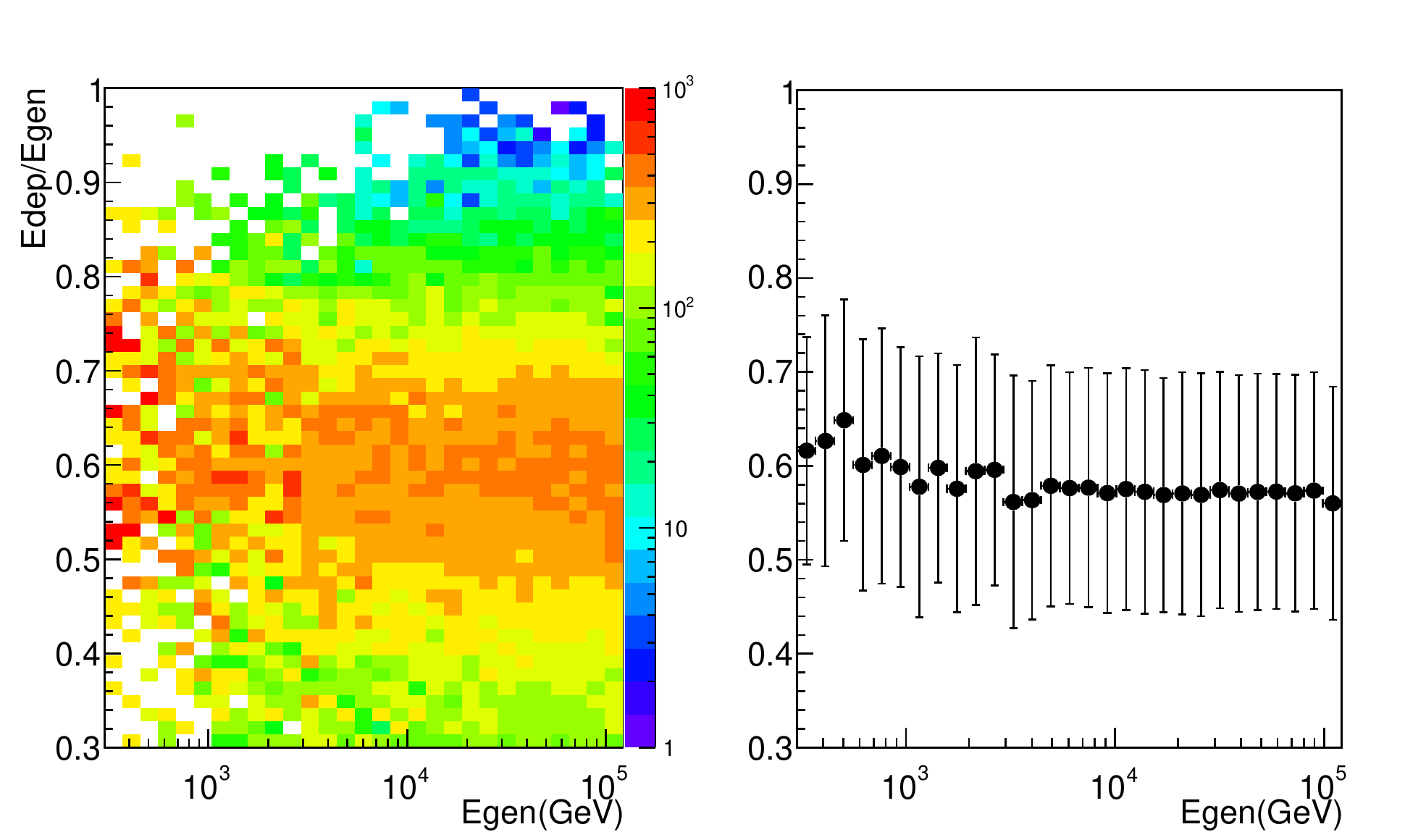}
\end{tabular}
\end{center}
\caption[example]
{ \label{fig:p_reco}
The ratios between energy deposition and incident energy for protons. The right panel shows the binned results of the left one.}
\end{figure}

Most particles entering the calorimeter initiate showers. The secondary particles' energies are deposited in CALO, and then collected and measured. For an electromagnetic (EM) shower induced by a high energy electron or gamma-ray, the energy resolution $\sigma_{E}/E$ of CALO can be parameterized as:
\begin{equation}
\frac{\sigma_{E}}{E} = \frac{A}{\sqrt{E}}\oplus \frac{B}{E} \oplus {C},
\end{equation}
where $\oplus$ represents quadratic addition. The stochastic term $A$ represents statistical-related fluctuations, such as intrinsic shower fluctuations, fluorescence and photoelectron generation and propagation statistics. The term $B$ is due to electronic noise. The constant contributions to the systematic term $C$ are detector calibration uncertainty and non-uniformity. As shown in Fig.~\ref{fig:e_reso}, the theoretical energy resolution of EM shower is about 0.1\% at 200 GeV. By considering the stochastic contribution with 10 photoelectrons per minimum ionization particle response, the resolution is increased to 0.45\%. With careful calibration, we believe the energy resolution for electrons and gamma-rays can be achieved below 1\%.

Hadronic showers produced by CR nuclei are considerably different from EM showers. The physical processes that determine the hadronic showers are: hadron production, nuclear de-excitation and charged hadron decay. An important characteristic of hadronic showers is that it takes longer to develop than the EM shower. Due to the finite volume of CALO, energy leakage of hadronic shower is inevitable. If early starting hadronic showers are selected with an efficiency of 50\%, as shown in Fig.~\ref{fig:p_reco}, the ratio between measured energy and incident energy is a constant, about 57\% above 1 TeV. The main contribution to the energy uncertainty is the intrinsic fluctuations of the hadronic interaction. Due to the deep nuclear interaction length of CALO, the energy resolution of protons can be confined to around 20\%, almost a constant from hundreds of GeV up to PeV.

\subsection{Particle Identification}
\label{sec:pid}
Primary CR consists largely of protons and helium nuclei, whereas its leptonic components accounts for only $10^{-2}$ to $10^{-3}$ of the total flux. Therefore, the main problem encountered in the spectrum measurement of electron is to suppress the proton background (e/p separation). CALO can be used not only for energy measurement but also for particle identification. Since the fact that EM and hadronic showers differ in their spatial and energy distributions in CALO, nine variables related to the longitudinal and traverse shower profiles are defined to solve the electron and proton identifying problem. 

\begin{figure}
\begin{minipage}[t]{0.5\linewidth}
\centering
\includegraphics[height=6cm]{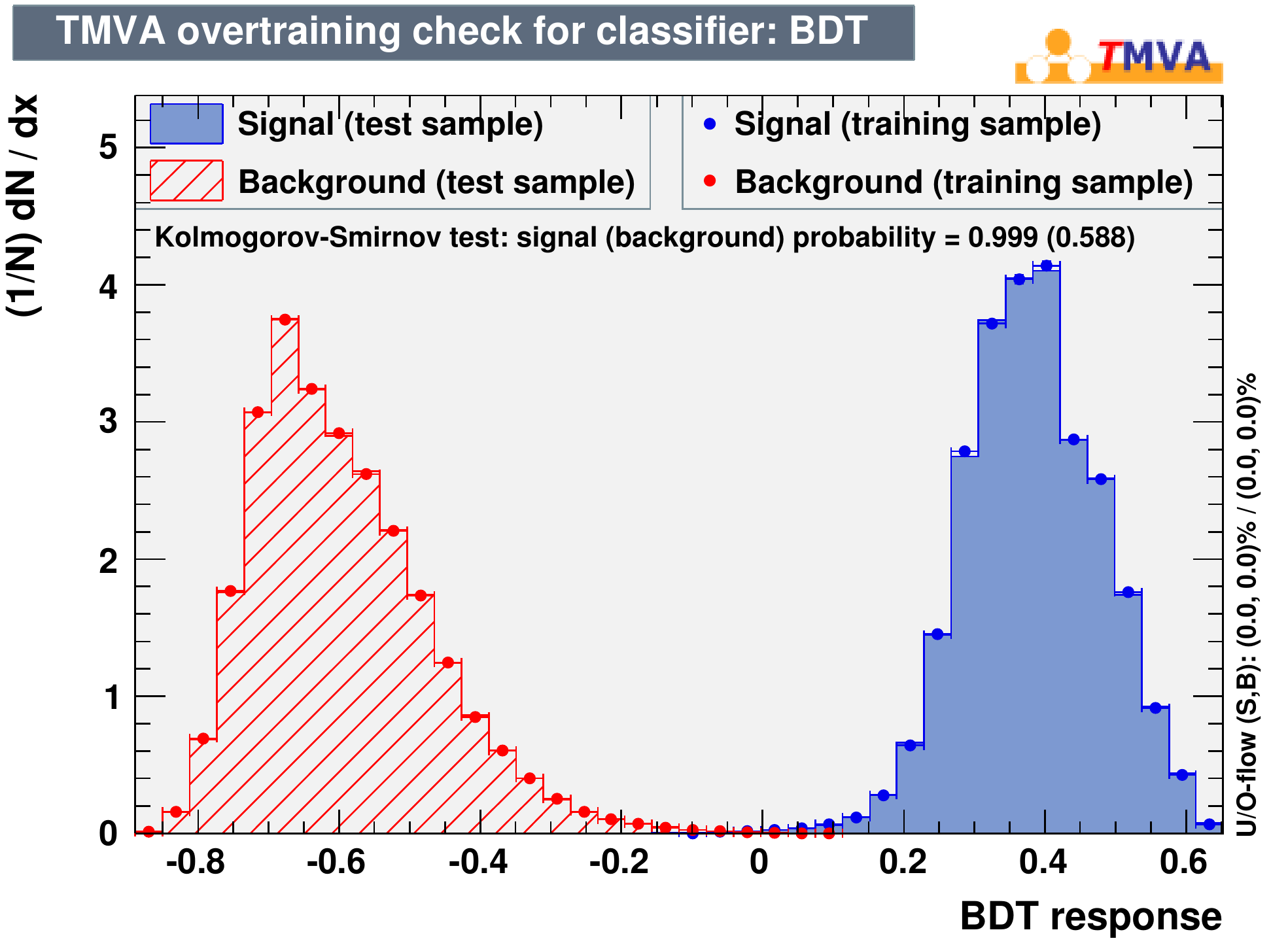}
\caption{BDT response for signal and backgournd.}
\label{fig:bdt}
\end{minipage}
\begin{minipage}[t]{0.5\linewidth}
\centering
\includegraphics[height=6cm]{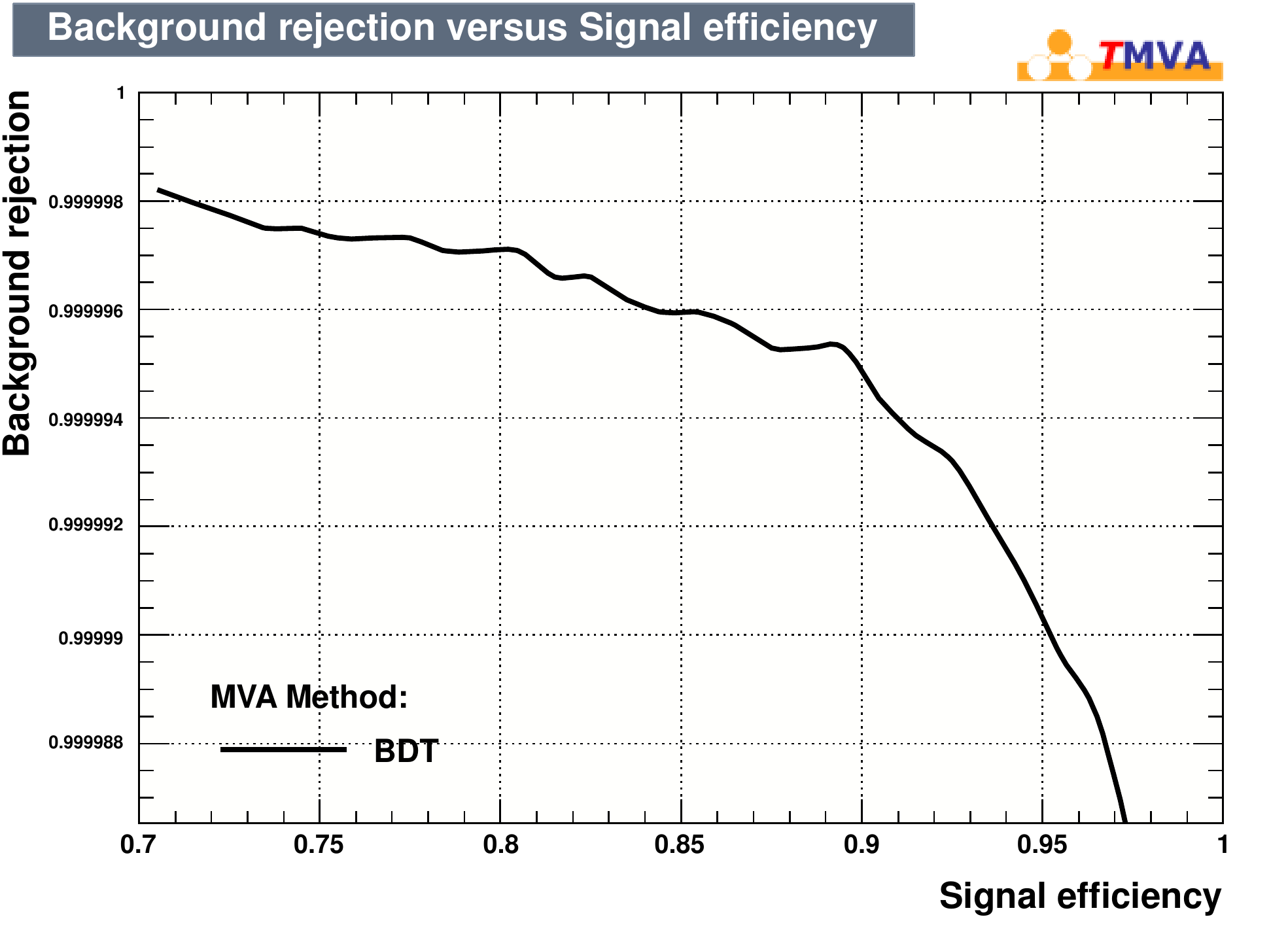}
\caption{Background rejection versus signal efficiency.}
\label{fig:pid}
\end{minipage}
\end{figure}

TMVA is a framework for training, testing and performance evaluation of multivariate classification techniques\cite{tmva}. It works in transparent factory mode to guarantee an unbiased performance comparison between the classifiers. The boosted/bagged decision trees (BDT) MVA method is selected for evaluating the CALO e/p separation performance. Fig.~\ref{fig:bdt} shows the machine learning BDT response for signals and background. Both signal efficiency and background rejection depend on the BDT cut value, as shown in Fig.~\ref{fig:pid}, the background efficiency is about $5\times10^{-6}$ when 90\% signals remain.

\subsection{Effective Geometrical Factor}
\label{sec:gf}

Effective geometrical factor connects the measurements and the physical variables,
\begin{equation}
N_{\rm obs} = F \times G_{\rm eff} \times T_{\rm exp} \times \eta_{\rm pid} \times \eta_{\rm reco},
\end{equation}
where $N_{\rm obs}$ is the number of observed events; $F$ is the absolute flux of the particle; $G_{\rm eff}$ represents the effective geometrical factor; $T_{\rm exp}$ means the exposure time; $\eta_{\rm pid}$ and $\eta_{\rm reco}$ represents particle identification and reconstruction efficiency. The effective geometrical factor can be calculated by a toy Monte Carlo method,
\begin{equation}
G_{\rm eff} = G_{\rm gen} \times \frac{N_{\rm pass}}{N_{\rm gen}},\end{equation}
for example, the geometrical factor $G_{\rm gen}$ of a flat panel receiving particles from only one side equals to $\pi S$, where $S$ represents the area of the panel. The effective geometrical factor is then proportional to the counting ratio between the panel $N_{\rm gen}$ and CALO $N_{\rm pass}$. The selection criteria $N_{\rm pass}$ in CALO are: 1). incident without earth and its atmosphere blocking; 2). CALO contains the shower maximum.

\begin{figure}
\begin{center}
\begin{tabular}{c}
\includegraphics[height=7cm]{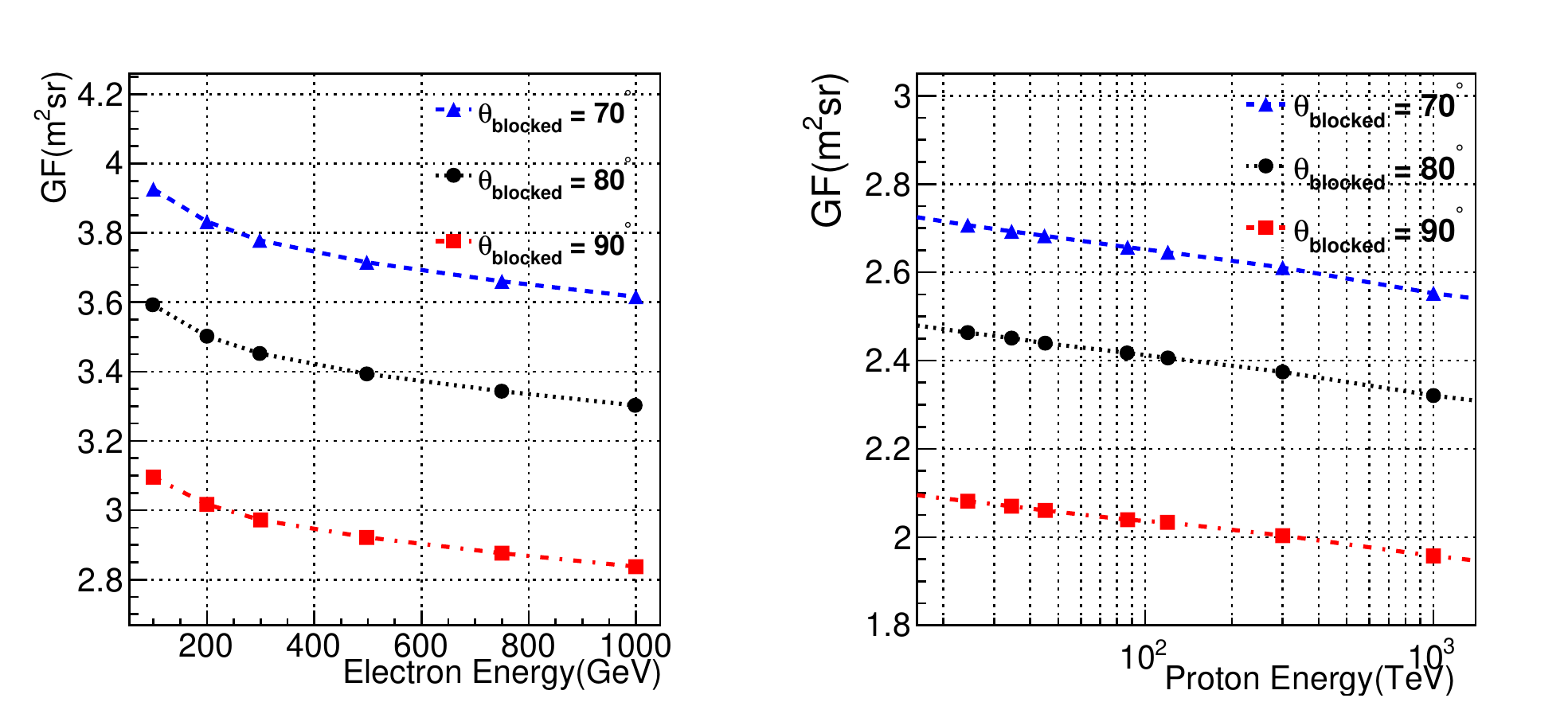}
\end{tabular}
\end{center}
\caption[example]
{ \label{fig:gf}
Effective geometrical factors of electrons and protons, respectively. Different block effects are considered to evaluate the CALO effective geometrical factors. ${\rm\theta}_{\rm {blocked}}$ = 70$^{\circ}$ represents 100 km atmosphere shielding effect. In the worst case that particles coming from the downstream are totally blocked by the platform (${\rm\theta}_{\rm {blocked}}$ = 90$^{\circ}$), the effective geometrical factor will be no less than 2.8 ${\rm m}^{2}{\rm sr}$ and 1.8 ${\rm m}^{2}{\rm sr}$ for electrons and CR nuclei, respectively.}
\end{figure}

Suppose HERD will be placed in 400 km LEO. Typically, the thickness of the atmosphere is about 100 km, which means the blocked angle is about 70$^{\circ}$. In addition to the shielding effect of the earth and atmosphere, some of the particles coming from the downstream will be blocked by the space station. The most extreme case is that events coming from the downstream are totally blocked by the platform. This worst case can be simply equivalent to the 90$^{\circ}$ blocked angle. Fig.~\ref{fig:gf} shows the effective geometrical factor for electrons and protons under different shielding effect, respectively. On average, the effective geometrical factor can be larger than 3 ${\rm m}^{2}{\rm sr}$ for electrons and 2 ${\rm m}^{2}{\rm sr}$ for protons.

\section{Expected performance of HERD}
\begin{figure}
\begin{center}
\begin{tabular}{c}
\includegraphics[height=8cm]{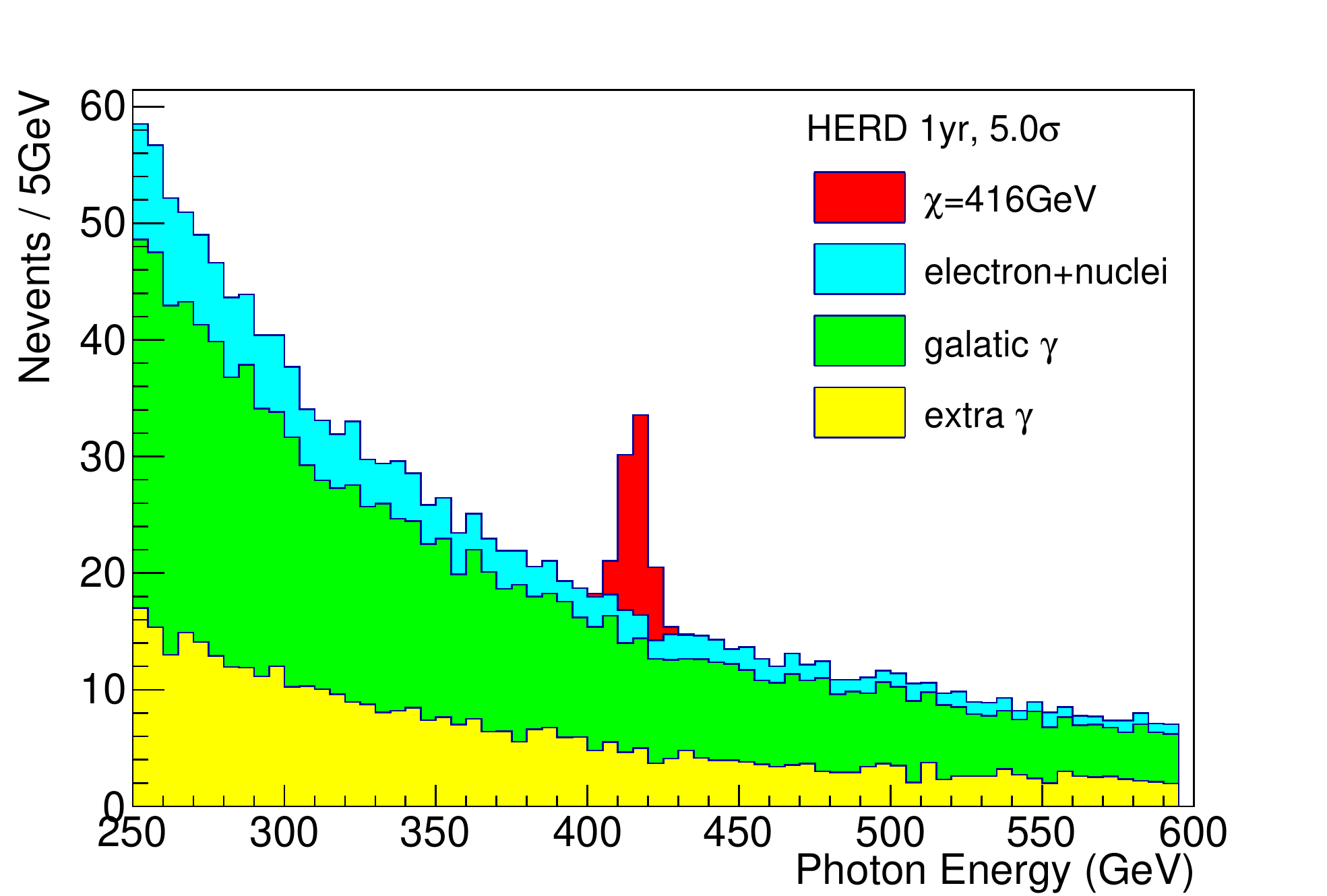}
\end{tabular}
\end{center}
\caption[example]
{ \label{fig:dm}
An example of HERD one-year observed photon smoking gun with a 5$\sigma$ significance, for DM mass of 416 GeV.}
\end{figure}

Suppose the DM annihilations can contribute to the diffuse gamma-ray flux through $\chi\chi \to \gamma\gamma$ and $\chi\chi \to \gamma Z_{0}$ channels. The kinetic energy of DM particles can be neglected since its present-day movement is non-relativistic. Therefore, the monochromatic gamma-rays are thought to be the smoking gun signal for identifying DM annihilation. The backgrounds for DM-induced monochromatic gamma-rays include CR nuclei, electrons and continuous gamma-rays. In this paper, we use a best fit formula according to many measurements\cite{gaisser} to represents the CR nuclei flux; for the electron spectrum, we adopt a broken power law parameterizaztion according to the measurements by Fermi\cite{Fermi_electron} and HESS\cite{HESS_electron}; for the continuous gamma-ray, including diffuse Galactic and extragalactic emission, we use the measurements by Fermi\cite{Fermi_gala,Fermi_extra}. The energy reconstruction is simply implemented by a Gaussian smear (CALO energy resolution) of the injection energy. The proton efficiency is $5 \times 10^{-6}$, as mentioned in Sec.~\ref{sec:pid}. Suppose TK can provide the charged/neutral particle separation power at least $10^{-3}$, here the charged particles efficiency is a fixed pass rate $10^{-3}$ when 90\% signals remain. Fig.~\ref{fig:dm} shows an example of reconstructed energy spectrum for the DM mass of 416 GeV, which corresponds to a minimum gamma-ray flux of $2.8\times10^{-10}$ ${\rm cm}^{-2}{\rm s}^{-1}$.

\begin{figure}
\begin{center}
\begin{tabular}{c}
\includegraphics[width=14cm,height=11cm]{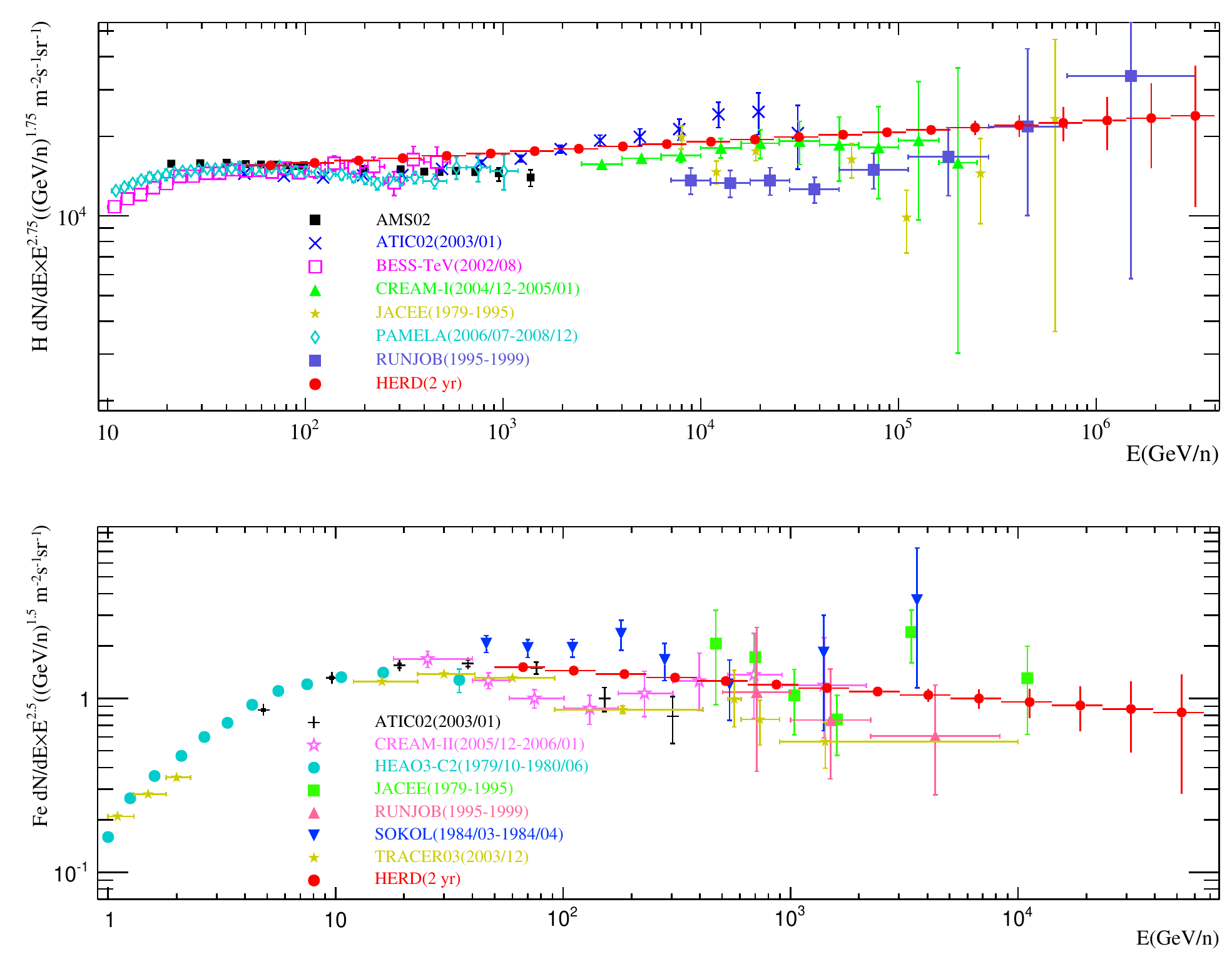}
\end{tabular}
\end{center}
\caption[example]
{ \label{fig:cr}
Expected results of proton and iron spectra from a two-year HERD observation based on the H{\H o}randel model\cite{horandel_model}, compared with previous measurements. Note that the flux is multiplied by $E^{2.75}$ and $E^{2.5}$ respectively to flatten the spectra and the error bars represent only the statistical uncertainties.}
\end{figure}

The expected results of protons and iron nuclei spectra from a two-year HERD observation are shown in Fig.~\ref{fig:cr}, compared with the previous experimental data including AMS02\cite{ams02}, ATIC-2\cite{atic2}, BESS\cite{bess}, CREAM,\cite{cream1,cream2} HEAO\cite{heao}, JACEE\cite{jacee}, PAMELA\cite{pamela}, RUNJOB\cite{runjob}, SOKOL\cite{sokol}, TRACER\cite{tracer}. HERD has the ability of precise CR spectrum and composition measurements up to the knee energy. At least ten events will be recorded from 900 TeV to 2 PeV for each of the chemical components, which means that the energy spectra of most nuclei will be directly extended to the knee range with smaller error bars than previous experiments.

\section{Summary}
Monte Carlo simulations show that HERD can achieve large effective geometrical factors ($>$3 ${\rm m}^{2}{\rm sr}$ for electrons and diffuse gamma-rays, $>$2 ${\rm m}^{2}{\rm sr}$ for CR nuclei) with high energy resolutions (1\%  for electrons and gamma-rays beyond 100 GeV, 20\% for protons from 100 GeV to 1 PeV) and e/p separation power better than $10^{-5}$. With these advantages, HERD will be a powerful instrument in space after 2020 for direct CR measurement, indirect search for DM, as well as for resolving the puzzles in the knee region of the CR spectrum and the anomalies in the electron spectrum.

\acknowledgments     
This work was supported by National Natural Science Foundation of China, Grant No.11327303; Cross-disciplinary Collaborative Teams Program for Science, Technology and Innovation, Chinese Academy of Sciences (Research Team of The High Energy cosmic-Radiation Detection); the Qianren start-up Program, Grant No.292012312D1117210; the Strategic Pioneer Program in Space Science, Chinese Academy of Sciences, Grant No.XDA04075600. 

\bibliography{report}   
\bibliographystyle{spiebib}   

\end{document}